\makeatletter
\@ifundefined{@parse@version@dash}{%
	\def\@parse@version#1{\@parse@version@0#1}
	\def\@parse@version@#1/#2/#3#4#5\@nil{%
		\@parse@version@dash#1-#2-#3#4\@nil}
	\def\@parse@version@dash#1-#2-#3#4#5\@nil{%
		\if\relax#2\relax\else#1\fi#2#3#4 }
}{}
\makeatother

\documentclass[%
reprint,
superscriptaddress,
amsmath,amssymb,
aps,
prl,
english,
]{revtex4-2}

\usepackage{graphicx}% Include figure files
\usepackage{dcolumn}% Align table columns on decimal point
\usepackage{bm}% bold math
\usepackage{float}
\usepackage[bookmarks=true,colorlinks,linkcolor=blue,urlcolor=blue,citecolor=blue]{hyperref}
\usepackage[mathlines]{lineno}% Enable numbering of text and display math
\newcommand{\mysection}[1]{\textbf{#1}}
\begin{document}
	
	\preprint{preprintnr}
	
	\title{Violation of the Wiedemann-Franz law in the Topological Kondo model}
	\author{Francesco Buccheri} \email{buccheri@hhu.de}
	\affiliation{Institut f\"ur Theoretische Physik, Heinrich-Heine Universit\"at, D-40225 D\"usseldorf, Germany}
	\author{Andrea Nava}% \email{andrea_nava_@msn.com}
	\affiliation{Dipartimento di Fisica, Universit\`a della Calabria, Arcavacata di Rende I-87036, Cosenza, Italy} 
	\affiliation{I.N.F.N. - Gruppo collegato di Cosenza, Arcavacata di Rende I-87036, Cosenza, Italy}
	\author{Reinhold Egger} % \email{egger@hhu.de}
	\affiliation{Institut f\"ur Theoretische Physik, Heinrich-Heine Universit\"at, D-40225 D\"usseldorf, Germany}
	\author{Pasquale Sodano}% \email{pasquale.sodano01@gmail.com}
	\affiliation{I.N.F.N., Sezione di Perugia, Via A. Pascoli, I-06123, Perugia, Italy}
	\author{Domenico Giuliano}  \email{domenico.giuliano@fis.unical.it}
	\affiliation{Dipartimento di Fisica, Universit\`a della Calabria, Arcavacata di Rende I-87036, Cosenza, Italy} 
	\affiliation{I.N.F.N. - Gruppo collegato di Cosenza, Arcavacata di Rende I-87036, Cosenza, Italy}
	
	
	\date{\today}
	
	\begin{abstract}
		We study the thermal transport through a Majorana island connected to multiple external quantum wires. In the presence of a large charging energy, we find that the Wiedemann-Franz law is nontrivially 
		violated at low temperature, contrarily to what happens for the overscreened Kondo effect and for nontopological junctions. For three wires, we find that the Lorenz ratio is rescaled by a universal factor $2/3$ 
		and we show that this behavior is due to the presence of localized Majorana modes on the island.
		%\begin{description}
		%\end{description}
	\end{abstract}
	
	\maketitle
	
	%\tableofcontents
	
	\mysection{\label{sec:Intro} Majorana fermions and the Wiedemann-Franz law.}
	Condensed matter realizations of Majorana fermions have been the object of active and
	continued interest during the last decade. Proposed platforms involve chains of magnetic adatoms, semiconductor-superconductor heterostructures in one and two dimensions and topological insulator-superconductor hybrid structures 
	\cite{Volovik1999,Kitaev2001,Fu2008,Lutchyn2010,Oreg2010,NadjPerge2013,Thakurathi2017,Hell2017}.
	Nonlocal transport, nontrivial braiding properties and intrinsic topological protection from disorder 
	effects have motivated the effort of embedding Majorana modes in an architecture for topological
	quantum computation \cite{Hyart2013,Plugge2016,Lutchyn2018,Sau2021}.
	%Intense experimental effort has provided clues about the possibility of realizing such physics in laboratory. 
	Experiments have insofar focused on the measurement of the local density of states %at the nanowire ends
	and on the charge transport properties of the Majorana quasiparticles, which include the detection of the quantized zero-bias 
	peak and of single-electron transport across a Coulomb-blockaded island \cite{NadjPerge2014,Wang2018,Whiticar2020}.
	While the existence of localized subgap states in several platforms has been confirmed,  most experimental observations 
	have not yet reached a sufficient accuracy or the observed properties could be otherwise explained, thus, their
	identification with Majorana zero modes (MZMs) has not been established beyond doubt. The lack of a clear-cut experimental evidence of MZMs  calls for a deeper understanding of a broader set of phenomena
	in superconductor-semiconductor systems \cite{Kells2012,Liu2012,Liu2017,Vuik2019,Chen2019}.
	\begin{figure}[b]
		\centering
		\includegraphics[width=0.45\textwidth]{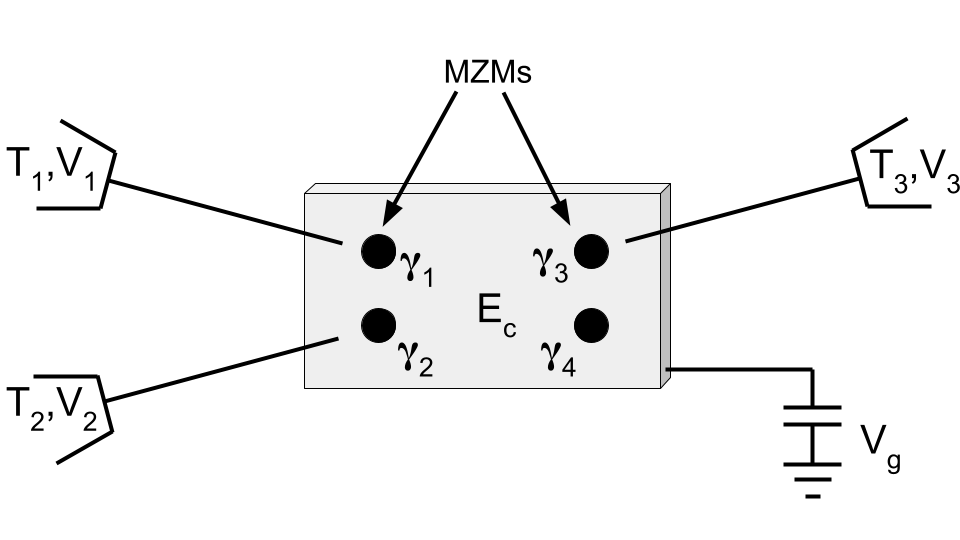}
		\caption{The Majorana-Coulomb box: a mesoscopic island hosting (an even number of) localized Majorana modes is connected to three external quantum wires as represented. At the far ends of these, external reservoirs control the distribution of the electrons injected into the junction.
			The scheme of this device is independent from the specific platform hosting Majorana quasiparticles.}
		\label{fig:system}
	\end{figure}
	\newline
	In this work, we study instead the heat transport across a junction hosting localized MZMs 
	\cite{Akhmerov2011,Lutchyn2017}, which is expected to provide a reliable signature of the topological quasiparticles \cite{Pan2021}, and look for a violation of the Wiedemann-Franz law (WFL) due to their presence.
	Typically, whenever charge and energy are carried by the same excitation(s), 
	a simple relation holds between the charge ($G$) and the thermal ($K$) conductance of the system at low temperatures: the WFL \cite{Franz1853,Benenti2017,Kumar1993}, stating that the Lorenz ratio \mbox{$L \equiv\frac{K}{TG}$} assumes the value \mbox{$L_0=\frac{\pi^2k_B^2}{3e^2}\approx 2.44\times 10^{-8}  W\Omega K^{-2}$}. 
	Electron-electron scattering or inelastic scattering processes (e.g., from electron-phonon interaction) can originate a violation of the WFL 
	\cite{Kane1992,Lavasani2019}. Yet, the remarkable renormalization of $L$ in interacting quantum wires \cite{Kane1996,Kane1997} 
	is expected to be washed out as soon as the wire is connected to  ideal Fermi liquid reservoirs, as in actual two-point charge/thermal 
	transport measurement \cite{Maslov1995,Oshikawa2006}.  In addition, in single-electrons transistors and quantum dots realized at the junction, 
	charging effects and resonant processes can determine $L\neq L_0$  \cite{Kubala2008,Karki2020c,Majidi2021}, by a factor which depends
	on the regime and the parameters of the model. However, as soon as the dot enters the Kondo regime, the emerging electron-electron 
	correlations restore the ideal value $L_0$ \cite{Boese2001,Costi2010}. Even when the impurity is overscreened in the multi-channel Kondo model, the very fact that charge and heat are still carried by the same excitation preserves the WFL \cite{Karki2020,vanDalum2020}.
	In the specific context of a multi-lead junction, we show that the MZMs  induce multiparticle resonant Andreev reflection and 
	crossed Andreev reflection processes at the Fermi level. Due to the fact that outgoing (from the junction) particles and 
	holes move in the same directions, but with opposite charges,  the corresponding contributions to the charge conductance differ from the one to the thermal conductance, thus inducing a renormalization of $L$. Based on this observation, in this  paper we relate the violation of the WFL and the universal renormalization factor acquired by $L$ to the appearance of the MZMs at the island and to the peculiar nature of the Topological Kondo fixed point.
	
	\mysection{ \label{sec:TK} The Topological Kondo effect.}
	We consider the system in figure \ref{fig:system}, in which a superconducting island hosting noninteracting MZM is connected to three quasi-1d interacting quantum wires. 
	The island is floating, subject to a constant gate voltage potential $V_g$, and with a large charging energy $E_c=e^2/2C$, 
	where $C$ is the total capacitance of the island.
	The Majorana modes $\gamma_a$ satisfy $\gamma_a = \gamma_a^\dagger$ and the Clifford algebra $\{ \gamma_a,\gamma_b \} = 2\delta_{ab}$.
	In the absence of other degrees of freedom with energy below the superconducting gap 
	$\Delta$, the fermion number parity of the island is conserved and, in each parity sector, the MZMs collectively encode a spin one-half degree of freedom. The quantum wires are described by the Tomonaga-Luttinger Hamiltonian 
	\begin{equation}
		H_0 =  \frac{u}{2}\sum_{a=1}^3 \: \intop_0^\ell dx 
		\left[g \left(\partial_x \phi_a\right)^2 + g^{-1} \left(\partial_x \theta_a\right)^2\right]\label{hlead}\;,
	\end{equation}
	\noindent
	with $\phi_a$ being the collective plasmon field of lead-$a$, $\theta_a$ its dual,
	$u$  the plasmon velocity and $g$ the Luttinger parameter ($g=1$ for non-interacting fermions). We set $\hbar=1$ throughout the paper. The wire length $\ell$ works  as a large-distance regulator and we eventually send it to infinity. 
	
	Starting from open boundary conditions on the quantum wires, each of them is connected at one end to one MZM via the term \cite{Fu2010,Herviou2016}
	\begin{equation}\label{Ht}
		H_{t}= - i\sum_{a=1}^3 t_{a} \gamma_{a} e^{i \chi} \Gamma_a e^{i\sqrt{\pi}\phi_a(0)} +h.c. \;,
	\end{equation}
	where the $t_a$ are the corresponding tunneling amplitudes and the operator $e^{2i \chi}$ creates a Cooper pair on the island.
	When bosonizing multiple wires, the Klein factors $\Gamma_a$ must be introduced \cite{Crampe2013,Tsvelik2013} in order to ensure the correct anticommutation relations. 
	Remarkably, it is possible to define a single complex fermionic degree of freedom by hybridizing the Klein factors with the localized MZMs, with occupation number $s_a=i\Gamma_a \gamma_a=\pm 1$  factored out of the 
	dynamics \cite{Beri2012}. As a consequence, the system admits a purely bosonic description, in contrast to the "non-topological" 
	junctions studied in \cite{Oshikawa2006,Hou2012}. 
	In the Coulomb-blockaded regime, cotunneling processes
	%on time scales $O(1/E_c)$,
	dominate the low-temperature transport. The effective Hamiltonian describing this physics is  $H =H_0 +H_K$, with   \cite{Beri2012,Beri2013,Altland2013} 
	\begin{equation}\label{Hjunction}
		H_K = - 2 J_K \sum_{a=1}^3
		\cos \left( \sqrt{2 \pi} \hat{k}_a \cdot \vec{\xi} ( 0 ) \right) .
	\end{equation}
	\noindent
	Here $J_K\sim t_a^2/E_c$ (anisotropy in the $J_K$ coupling for different legs is irrelevant under renormalization \cite{Beri2012,Altland2013}),  $\hat{k}_a=\left(\cos\frac{4\pi (a-1)}{3},\sin\frac{4\pi (a-1)}{3}\right)$ and the 
	relative fields are $\xi_1 ( x ) = \frac{\phi_1 ( x ) - \phi_2 ( x ) }{\sqrt{2}}, 
	\xi_2 ( x ) = \frac{\phi_1 ( x ) + \phi_2 ( x ) - 2 \phi_3 ( x ) }{\sqrt{6}}$.
	The boundary interaction \eqref{Hjunction} emerges in a number of different models,
	such as the planar quantum Brownian motion in a periodic potential \cite{Yi1998}, or a 
	junction of three Bose liquids, or of Josephson junction chains \cite{Tokuno2008,Giuliano2010}. 
	In a grounded island, the strong hybridization of the wires with the MZMs at low temperatures enforces Andreev reflection around the Fermi energy \cite{Altland2013}, i.e., the boundary condition $\phi_{a}(0)=0$ on each wire.  When, instead, the island is floating as in the TK
	model, the constraint of charge conservation implies that the total charge mode $\sum_{a=1}^3 \phi_a$ satisfies Neumann boundary conditions at the origin and Andreev boundary conditions can only be imposed on the relative fields $\xi_{1,2}$. The dominant processes at low
	temperatures are correlated crossed Andreev reflections among different wires, \cite{Beri2012,Altland2013}, resulting in the net transmission of an effective fractionalized $2e/3$ charge \cite{Nayak1997}.  Indeed, for \mbox{$g>3/4$}, the system possesses a stable strong-coupling 
	fixed point (FP), isotropic with respect to the leg indexes, the "topological Kondo" (TK) FP. For $3/4<g<1$, both weak- and strong-coupling FPs are stable and a first-order transition in the boundary coupling strength appears \cite{Herviou2016}. The charge conductance, together 
	with the thermodynamic properties, identifies the TK FP as a non-Fermi liquid \cite{Beri2012,Altland2013,Altland2014,Galpin2014,Buccheri2015}. 
	The crossover temperature for $g=1$
	\begin{equation}\label{TK}
		k_B T_K \sim E_c e^{-1/(2\rho_0 J_K)},
	\end{equation}
	where $\rho_0$ is the density of states at the Fermi energy, represents the scale for the onset of the Kondo correlations.
	%As we discuss in the following, the charge conductance, together with the thermal conductance properties, identify the TK FP as a non-Fermi liquid hosting a remarkable breaking of the WFL. 
	
	\mysection{\label{sec:WFL} Electric and thermal conductances and WFL at weak coupling.}
	We connect the quantum wires to Fermi liquid reservoirs, described by the corresponding Fermi distributions $f_a ( E ) = [1+e^{(E-\mu_a)/k_BT_a}]^{-1}$, with all the temperatures set below $T_K$.
	%The voltage biases on each reservoir are tuned in such a
	%way that the system is at charge neutrality, while the temperatures are all set below $T_K$.
	%In order to be able to tune the $\mu_a$ and the $T_a$, 
	We assume that the reservoirs all have large, though finite, charge and thermal capacitances. Thus, we may change the  distributions of electrons entering the junction from reservoir $a$ by varying $\mu_a$ and/or $T_a$ by  $ e \Delta V_a$ and/or $\Delta T_a$ around common reference values $\mu,T$.
	Setting nonzero biases, we induce  a net charge and heat current flow into the system, 
	carried by the ballistic Luttinger  liquid excitations. 
	Thermal current measurements can be performed by detecting the temperature gradient across a contact between the wires and the reservoir, or the rate of change of the reservoir temperature \cite{Giazotto2006,Benenti2017,Pekola2021}. %$I_{h,a}=\left(C_a +\partial_T C_a\right) \dot{T_a}$. 
	The charge and the thermal currents in lead $a$, $I_{e,a},I_{h,a}$, are related to $\Delta V_b , \Delta T_b$ 
	via the respective conductances \cite{Benenti2017}. Within linear response, these 
	are defined as
	\begin{equation} \label{conductances}
		G_{a,b}=\left(\frac{I_{e,a}}{ \Delta V_b}\right)_{\Delta T_b = 0}
		\qquad 
		K_{a,b}=\left(\frac{I_{h,a}}{\Delta T_b}\right)_{I_{e,b} = 0}
	\end{equation}
	We consider here small variations of the chemical potential around charge neutrality: under these conditions, the Seebeck and the Peltier coefficients vanish, due to particle-hole symmetry.
	
\begin{figure}
		\centering
		\includegraphics[height=0.19\textheight]{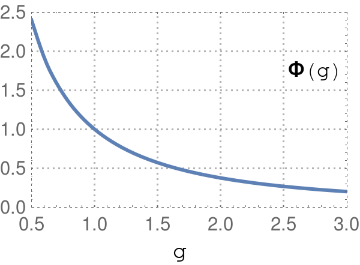}
		\caption{  $\Phi(g)=L/L_0$ \eqref{Phi} at weak coupling as a function of 
			$g$ for $0.5 \leq g \leq 3$. By exact calculation, we find $\Phi (g=1)=1$.}
		\label{fig:Phi}
	\end{figure}

	The conductances \eqref{conductances} are conveniently computed by means of the chiral fields
	\mbox{$2\varphi_{a,R/L} ( x   ) =g^{1/2}\phi_a ( x   ) \pm g^{-1/2}\theta_a ( x   )$}.
	The high- and low-temperature FPs are described 
	by linear relations between such chiral fields, in the form 
	\mbox{$\varphi_{a,R}( x  )=\sum_{b=1}^3 \: \rho_{a,b}\varphi_{b,L}(-x  )$}, 
	with \mbox{$\rho_{a,b} = \delta_{a,b}$} for open boundary conditions and
	\mbox{$\rho_{a,b} = \frac{2}{3} -\delta_{a,b}$} at the TK FP \cite{Oshikawa2006,Eriksson2014}.
	Accordingly, we introduce the ``unfolded'', left-handed fields
	\mbox{$\varphi_a ( x ) = \varphi_{a,L} ( x ) H ( x ) + \sum_{b=1}^3 \rho_{b,a} \varphi_{ b,R} (-x ) H ( - x )$}, with $H(x)$ being the Heaviside's step function.
	The fields $\varphi_a ( x )$ are continuous across the junction: consistently with \cite{Kane1996,Kane1997}, 
	we assume that $\varphi_a ( x )$ is injected from reservoir $a$ into the corresponding wire at $x=\ell$ and propagates toward the junction by always keeping at chemical and thermal equilibrium with that reservoir. 
	In the presence of a voltage bias $ V_a $, we rewrite the fixed point Hamiltonian in terms 
	of the $\varphi_a$ as
	\begin{equation}\label{HaVa}
		H_{\rm FP} = \sum_{a=1}^3 \: \intop_{-\ell}^{\ell} dx \left\{ u\left( \partial_x \varphi_a ( x ) \right)^2
		+ e \sqrt{\frac{g}{\pi}} V_a \partial_x \varphi_a ( x ) \right\}\;,
	\end{equation}
	\noindent
	where $e \sqrt{\frac{g}{\pi}} \partial_x \varphi_a ( x )$ is the charge density for 
	the chiral field $\varphi_a ( x )$.
	The corresponding charge and thermal current densities are
	\begin{eqnarray}
		j_{e,a} ( x ) &=& eu\sqrt{\frac{g}{\pi}}\left\{ \sum_{b=1}^3 \rho_{b,a} \partial_x \varphi_b ( -x) - \partial_x\varphi_a(x) \right\} \nonumber \\
		j_{h,a} ( x ) &=& u^2 \left\{ ( \sum_{b=1}^3 \rho_{b,a} \partial_x \varphi_b ( - x ))^2 - ( \partial_x \varphi_a (x ))^2 \right\} . \label{currentop}
	\end{eqnarray}
	The effect of the potential is reabsorbed via the shift of the bosonic field $\partial_x \varphi_a ( x ) \to \partial_x \varphi_a ( x ) + \frac{e V_a}{u} \sqrt{\frac{g}{\pi}}$.
	This induces a corresponding shift in the current density operators and in their expectation values.
	From now on, we consider the system at charge neutrality $V_a=0$, where the thermopower vanishes identically due to particle-hole symmetry. 
	The first non-vanishing off-diagonal contribution to the conductances at weak coupling arises at 
	second order in $J_K$ in \eqref{Hjunction}. 
	Within the linear response framework \cite{Campagnano2016,Us2021}, we obtain
	\begin{equation}\label{eq:G}
		G_{a , b} =  \frac{6 \pi e^2 \Gamma^2 \left( 1 / g \right)  }{  \Gamma   \left( 2 /g \right) }
		\tilde{J}^2_K \left(T\right)\left(\frac{1}{3}-\delta_{a,b}\right) \;,
	\end{equation}
	where the effective dimensionless coupling constant is \mbox{$\tilde{J}_K\left( T \right)=J_K E_c^{-\frac{1}{g}}\left( 2 \pi k_B T \right)^{\frac{1}{g}-1}$}. 
	The heat conductance is $K_{a , b }=L_0\Phi(g)G_{a , b} T$, with the Lorenz ratio expressed in terms of the dimensionless function
	\begin{eqnarray}  \label{Phi}
		\Phi ( g )&=&  \frac{3  \Gamma \left(  2/g\right)}{g \pi \Gamma^4 \left( 1/ g \right)} 
		\int \: d z \: d w \: \frac{w}{\sinh ( \pi z ) } \times 
		\\
		&&   \left| \Gamma \left( 1 / ( 2g )  +  i \left( z    - w \right) \right) \right|^2
		\left| \Gamma \left( 1 / ( 2 g ) +  i   w \right) \right|^2  , \nonumber 
	\end{eqnarray}
	in which $\Gamma$ denotes the Euler's Gamma function. This expression is plotted in fig. \ref{fig:Phi}.
	In the non-interacting case, or rather, for non-interacting reservoirs $g=1$ (see below), \eqref{Phi} can be computed exactly and the Lorenz ratio shown to be equal to $L_0$.
	The physical mechanism behind this relation is the fact that, at weak coupling $\tilde{J}_K\ll1$, charge and energy are carried by electrons that tunnel through the island. This picture ultimately breaks down at strong coupling, due to the tunneling of fractional-charge quasiparticles or, equivalently, to the onset of multiparticle Andreev and crossed Andreev reflections \cite{Beri2017,Ponomarenko1995,Safi1995,Nayak1997}.
	While an exact solution of the model has been provided in \cite{Altland2014b,Buccheri2015} for every value of $J_K$, we are not aware of results concerning the conductance in intermediate regimes.
	
	\mysection{\label{sec:noWFL} Violation of the WFL at strong coupling.}
	We now show that at the strong-coupling FP the Lorenz ratio differs from the ideal value by a factor, 
	which is composed of a universal part and of a simple function of the Luttinger parameter only. Given its purely bosonic description discussed above, the problem at hand can be tackled within 
	the general framework of \cite{Bellazzini2006,Bellazzini2007,Mintchev2013}.
	Averaging the shifted current operators \eqref{currentop} we obtain the charge and the thermal conductance tensors given by 
	\begin{equation}
		G_{a , b}   =  \frac{ e^2 g}{2 \pi}  \{ \rho_{ a, b  } -  \delta_{  a , b }  \} \;,\; 
		{ K}_{a , b}   = \frac{\pi  k_B^2 T}{6} \{ [ \rho_{  a,b   } ]^2 - \delta_{ a , b  } \} \;, \label{KTK}
	\end{equation}
	\noindent
	respectively. The tensor structure of the thermal conductance, written in terms of the splitting matrix of the plasmon excitations, highlights how the energy is divided between the leads at the low-temperature fixed point. We obtain the nontrivial result
	\begin{equation}\label{LRTK}
		\frac{ {K}_{a , b}}{T { G}_{a , b}}
		= \frac{L_0}{g}
		\left\{   \delta_{  a , b }  + \rho_{  a , b}  \right\} =\frac{2}{3}\frac{L_0}{g}\; ,
		\forall a,b  \: .
	\end{equation}
	Eq.\eqref{LRTK} evidences a remarkable breakdown of the WFL at temperatures below $T_K$ in \eqref{TK}. This is determined by a combination of the bulk interaction in the leads \cite{Kane1996} and of the onset of multiparticle scattering processes at the junction.  The former effect is nonuniversal (such as the Luttinger parameter $g$) and, more importantly, it is washed out once we connect the wires to external Fermi liquid reservoirs, to which the temperature and voltage biases are applied. To see this, we  picture the reservoirs as a discontinuity in the Luttinger parameter:  $g=g(x)=g H(\ell-x)+g_R H(x-\ell)$, with $g_R=1$. In this setting, the interaction parameter $g$ in \eqref{KTK} must be substituted with $g_R$. In particular, the ${\bf \rho}$ matrix is replaced by \cite{Maslov1995,Ponomarenko1995,Safi1995}
	\begin{equation}\label{hatrho}
		\rho\left(g\right) \to  \frac{\left(g_R   -   g\right)\mathbb{I}+ \left(g_R   + g 
			\right)\rho}{ \left(g_R  +  g\right)\mathbb{I}+ \left(g_R  -   g\right)\rho }\;=\;\hat{\rho}\left(g_R\right)\;.
	\end{equation}
	This identity can be directly verified by substituting a generic form of the $\rho$ matrix satisfying charge conservation \cite{Us2021}.
	\noindent
	%When \eqref{hatrho} (with $g_R=1$), rather than 
	%$\rho$,  into \eqref{LRTK}, we find that the 
	When the reservoirs supply quasiparticles with a free Fermi distribution, the only corrections to the Lorenz ratio arise from correlated multiparticle scattering processes at the junction, encoded in the $\rho$ matrix in \eqref{KTK}. 
	Moreover, when the junction can be described by means of a single-particle scattering matrix, the charge and the thermal conductance necessarily have the same tensor structure and the WFL is readily verified \cite{Benenti2017,Us2021}. 
	Hence, the Lorenz ratio \eqref{LRTK} can be regarded as a signature of strong correlations. 
	%{\color{blue} and the scattering processes  are junction are fully described by means of a single-particle \buc{scattering} matrix, the charge and the thermal conductance necessarily have the same tensor structure and the WFL is readily verified \cite{Benenti2017,Us2021}. }
	Corrections to the conductance tensors at the strongly-coupled FP are dictated by the dimension of the leading irrelevant operator. The heat conductance receives corrections that scale as $\sim T^{(8g-3)/3}$, while corrections to the charge conductance and the Lorenz ratio \eqref{LRTK} scale as $\sim T^{2(4g-3)/3}$.
	
	Let us compare this result with similar "non-topological" junctions. For instance, whenever the island is 
	short-circuited and the transport does not involve MZMs, the analysis falls back to the case of 
	\cite{Oshikawa2006}. In \cite{Us2021}, it is shown that no violation of the WFL occurs at low temperature. 
	It is also important to compare with the situation in which almost-zero-energy fermionic modes are present on the island, instead of MZMs. 
	Whenever the gate voltage is tuned so that a single electron or a single hole is present in the dots, the problem maps onto a $SU(3)$ Coqblin-Schrieffer model. In the ground state, the impurity degree of freedom on the island is exactly screened by the conduction electrons and the Lorenz ratio is equal to $L_0$, up to corrections that scale as $\left({T}/{T_K}\right)^2$ \cite{Mora2009,Carmi2011}.

	\mysection{\label{sec:Discussion} Discussion.}
	We have found a class of systems in which the WFL  is nontrivially violated, see also \cite{Us2021}. A value of the Lorenz ratio
	different from $L_0$  is generically expected in one and higher dimensions in the presence of electron-electron interactions \cite{Kane1996,Lucas2018}. However, while in 
	one-dimensional systems the effect of the interaction is fully encoded in the Luttinger parameter $g$, 
	we have shown that, when a junction between 1d Tomonaga Luttinger liquids is connected to Fermi liquid reservoirs,
	at low enough energies the reservoirs always renormalize the parameter $g$ back to the noninteracting limit, washing out the corresponding renormalization of the response functions and of the Lorenz ratio \cite{Maslov1995,Oshikawa2006}. 
	Yet, a nontrivial splitting matrix ${\bf \rho}$ at the junction can give rise to a renormalization of the Lorenz ratio (by a universal factor $\frac{2}{3}$), despite the junction being made out of effectively noninteracting leads. 
	%An intuitive explanation of our result is that \
	Indeed, the interplay between the Majorana-enforced boundary conditions and charge conservation implies that the low-temperature physics is described by the TK
	boundary conditions, enforcing that an injected particle of charge $e$ is equally "split" into fractional-charge- $\frac{2e}{3}$ particles propagating into the other two wires, and backscattered as a fractional, charge-$\frac{e}{3}$ hole. Particles and holes propagating in the same/opposite direction give opposite/equal sign contributions to the charge current and equal/opposite sign contributions to the thermal current: this originates the charge-heat separation, witnessed by the breakdown of the WFL at the TK FP. 
	The emergence of fractional quasiparticles in this system has been discussed in connection with the shot noise in \cite{Beri2017}.

	No violation of the WFL would be expected at low temperatures %at the free-fermion point. This holds true
	in the presence of weak impurity scattering \cite{Lavasani2019} and in the Kondo regime. While this is not surprising for the underscreened and the exactly screened Kondo effect, where the system flows toward a Fermi liquid FP \cite{Nozieres1974}, the WFL is verified also in the multichannel version \cite{Karki2020,vanDalum2020}, even though the Fermi liquid picture breaks down \cite{Affleck1991f,Affleck1991,Affleck1993}.
	We instead attribute the value of the Lorenz ratio found here to the fractionalization of the charge transport originated by the MZMs \cite{Beri2017}. 
	Indeed, while recovering a stable non-trivial FP typically requires a large interaction in the leads, the emergence of localized degrees of freedom triggers here a low-temperature evolution of the system toward a non-trivial FP \cite{Akhmerov2011,Fidkowski2012,Affleck2013,Pan2021}, 
	which would be otherwise unstable \cite{Oshikawa2006}. 
	The renormalization of the Lorenz ratio by the universal factor $2/3$ emerges as a hallmark of the presence of MZMs at the junction, independently of the specific platform. A generalization to multiple wires is discussed in \cite{Us2021}. A violation of the WFL in related systems has been reported numerically in \cite{RamosAndrade2016,Ricco2018}.
	
	As phonons carry heat, but not charge, they can affect the measured Lorenz ratio. In a single wire, phonons generate a quantized heat conductance \cite{Rego1998}, %with each phonon channel contributing with the universal value $\pi k_B^2 T/6$.
	but the contribution arising from tunneling of phonons in mesoscopic heterostructures depends on several microscopic details of the system 
	%of the interface, on the material and on the number of phonon modes
	\cite{Prunnila2010}. 
	In general, it does not have the characteristic tensor structure of \eqref{KTK} and \eqref{LRTK}, since phonons are not affected by the charging energy, which allows to separate their effects from electronic effects. 
	%A conductance of the order of the conductance quantum can happen only with specific settings
	Control of the heat carried by phonons is currently object of extensive studies \cite{Benenti2017}.%Yu1995
	
	The devices in \cite{Kanne2021,Vekris2021,Vekris2021x} potentially provide a candidate for observing the (TK) fixed point. %  \cite{Lutchyn2010,Oreg2010,Song2021}
	For instance, assuming an effective coupling in \eqref{Hjunction} of \mbox{$J_K \approx 0.02 meV$}, a density of states $ \rho_0\sim D_0^{-1}$ and a bandwidth \mbox{$D_0 \sim E_c \sim 0.2 meV$}, one obtains a Kondo temperature \mbox{$T_K \sim 16 mK$}. The off-diagonal thermal conductance at $T=1 mK$ is \mbox{$K_{12}\approx 3\times 10^4 \frac{eV}{K\,s}$}. 
	%The sensitivity of the Kondo temperature to the contact parameters and to the wire details poses a challenge to experiments.
	Another platform may be provided by the cold-atomic setup proposed in \cite{Buccheri2016}: for the realized trap, a thermal  conductance of \mbox{$K_{12}\approx 3 \frac{meV}{K\,s}$ at $T \sim T_K \sim 10nK$ is expected.} Temperature control and measurement with the required precision appears to be within reach of present-day techniques \cite{Giazotto2006,Majidi2021}.

	%In one-dimensonal nanowires, within the continuum approximation, %Yu1995

	% Many of the setups in which the emergence of Majorana modes (MZMs) has been theoretically predicted are theoretically described in terms of pertinent quantum impurity models, connected to one-dimensional (noninteracting or  interacting) leads, that is, junctions of quantum wires (JQWs) \cite{Lutchyn2010,Oreg2010,Fidkowski2012,Affleck2013}. Sharp features in, e.g., the transport properties of a JQW typically appear at nontrivial fixed points (NFP) of the corresponding phase diagram, which cannot be described only in terms of single-particle scattering processes at the junction \cite{Oshikawa2006,Hou2012}. Recovering a stable NFP typically requires an unphysically large interaction in the leads, or the emergence of localized degrees of freedom which allow for localized interactions triggering a low-energy/low-temperature evolution of the system toward a NFP. This last situation typically realized in junctions hosting localized MZMs \cite{Akhmerov2011,Fidkowski2012,Affleck2013,Pan2021}. 
	\vspace{0.06cm}
	
	\mysection{Acknowledgments}
	\textit{F.B. and R.E. were funded by the Deutsche Forschungsgemeinschaft (DFG, German Research Foundation) under Germany's Excellence Strategy Cluster of Excellence Matter and Light for Quantum Computing (ML4Q) \mbox{EXC 2004/1  390534769} and Normalverfahren Projektnummer \mbox{EG 96-13/1} and under Projektnummer \mbox{277101999 - TRR 183} (project B04).  
		A. N. was financially supported  by POR Calabria FESR-FSE 2014/2020 - Linea B) Azione 10.5.12,  grant no.~A.5.1.  D.G.   acknowledges  
		financial support  from Italy's MIUR  PRIN project  TOP-SPIN (Grant No. PRIN 20177SL7HC). We thank M. Burrello for the valuable feedback.
	}
	
	\bibliographystyle{apsrev4-1}
	\bibliography{YHbib}

\end{document}